# STATISTICAL FEATURES OF THE WIND FIELD OVER THE INDIAN OCEAN FOR THE PERIOD 1998-2008

Polnikov V. [1)], Pogarskii F.[1)], Sannasiraj S.[2)],

[1)] Obukhov Institute of Atmospheric Physics of RAS, Moscow, Russia

E-mail: polnikov@mail.ru

[2)] Indian Institute of Technology of Madras, Chennai, India

**Abstract**

We have done a statistical analysis of the wind field from the archive of NCEP / NOAA over the Indian Ocean for the period 1998-2008yy, which is given on a grid 1x1.25 of latitude-longitude with 3h time-step. Initial analysis includes mapping the average wind fields $<W(\mathbf{x},T)>$ and fields of mean density of the wind-kinetic-energy flux $<E_A(\mathbf{x},T)>$, obtained with different periods of time averaging $T$, as well as the assessment of 11-year trends in these fields.

The subsequent analysis is concerned with partition of the Indian ocean area into 6 zones, provided by the spatial inhomogeneity of the analyzed wind field. This analysis includes: (a) an assessment of temporal variations for the wind speed field averaged over the Ocean and the zones, $<W(\mathbf{x},T)>$, and for the field of wind-energy flux $<E_A(\mathbf{x},T)>$; (b) construction of time history series of these fields averaged with different scales and estimating frequency spectra of these series; (c) finding the extremes of the wind field (in the zones of Ocean); (d) construction of histograms of the wind field; and (e) calculation of the first four statistical moments for the wind field (in zones and in the whole Ocean). The results obtained allow us to estimate stored energy of the wind field in the Indian Ocean, its climate variability, and the distribution of the above statistical characteristics of wind field in the Indian Ocean.

**Introduction and statement of objectives**

This paper represents an initial phase of implementing an international Russian-Indian project devoted to a study of wind- and wave-field energy-store, processes of their interaction, and long-term variability of these fields on the example of the Indian Ocean (IO). Because the wind is the primary source of all the processes mentioned above, the study of its variability and statistical characteristics is of the primary interest and obligatory. In addition, we have developed and presented in this paper an approach to handling geophysical data, which will be used later as a reference for processing fields of other variables (the wind sea, the rate of energy transfer from wind to waves, etc.), planned for implementation of this project.

Wind field, used in this study, was obtained from the website [1] in the form of three-dimensional arrays for two wind components. These data represent a reanalysis made in the process of application of wind-wave numerical model WAVEWATCH (WW) [2]. Grid spacing is $1^0$ in latitude, $1.25^0$ in longitude, and 3h in time; the accuracy of the field components is of the order of 1 m/sec. A complete history of the specified wind field is beginning in 1997 and continuing up to the present. At the time of our work organization, the field was available to us for the period of 11 years: 1998-2008.

Such a detailed wind field, to our knowledge, has not been used for a detailed examination of its statistics in the Indian Ocean yet, despite the fact that the works of such a kind have a long history. They were successfully carried out by many researchers around the world (see references in [3-8]). Thus, among Russian authors we can mention the studies by Gulev et al [3], in which the issues of climate variability of the wave and wind fields were considered on the basis of voluntary ship observations, extrapolated to a much more greater space grid (eventually, $15^0 \times 15^0$). A wide range of issues were addressed in the group of Lapotukhin [4], based on reanalysis data for wind fields of various origins and on numerical simulation for waves. A wider range of issues is studied by scientists from other countries, based on the reanalysis data ERA-40, performed on the grid $1.5^0 \times 1.5^0$ for observation period starting from the end of 50s of last century [5]. At this level of studies, the analysis of wind fields is added already wave fields calculated on the basis of some numerical models (in Europe, mainly WAM [9], in other parts of the world, mostly WW [10]). And finally, the statistics and variability of wind fields has been extensively studied on the basis of satellite data [6-8]. In such cases, the grid density can reach a density of $0.25^0 \times 0.25^0$ with an average time step of 6 (or more) hours.

The objectives of our work does not include any detailed analysis of these works, as our project focuses primarily on the study of peculiarities of the space-time structure of wind and wave energy, as well as the variability of these fields and their interaction on the climate scale. Therefore, for our purposes, the above-noted works of others authors are only a certain frame permitting a control of the "reasonableness" of those of our results which are corresponding to with the above ones.

For further, it is important to note a detailed account of the spatial inhomogeneity of the wind field in the Indian Ocean, as a principal feature of our study. Here, this circumstance is solved by the partition the whole IO into a number of zones differing in the dynamics of wind field. In this case, an initial qualitative (visual) analysis of the wind is required, which is completed by a quantitative analysis of the averaged wind fields. On this basis, the partition into zones was carried out.

In this paper, the partition of Indian Ocean (IO) is realized by dividing the entire area into 6 zones:

1) North West (Z1) with the coordinates of boundaries (40E - 80E; 25N - 7N);
2) North East (Z2) - (80E - 100E; 25N - 7N);
3) Equatorial (Z3) - (35E - 105E; 7N - 9S);
4) South trade-wind (Z4) - (35E - 142E; 9S - 22S);
5) South Subtropical (Z5) - (20E - 140E; 22S - 35S);
6) South Indian (Z6) - (20E - 147E; 35N - 60N).

As an explanation, we note that the naming of the partition area as "the north, the equatorial, and so on" is quite obvious. In true, in this case, the "equatorial" does not coincide with the geographical equator, due to the spatial spreading the field pattern of the mean wind (see below).

Partition of the northern part of IO in the eastern and western areas is also obvious: they are separated the Indian peninsula significantly affecting the spatial dynamics of the wind. Equatorial zone has a specific circulation, and a partitioning the southern part of IO into three zones is due to the presence of South-trade wind zone of significant winds, alternated by the area of less regular and weaker winds, which would naturally be viewed separately from the rest of the (South)Indian Ocean. The most southern border is given by the conditionally-mean ice edge (which is more important for the wave field consideration).

The general formulation of the basic set of the tasks is as following.

1. To build and analyze four types of maps for the mean wind field $<W(i,j,T)>$ given by the ratio

$$<W(i,j,T)> = \left( \sum_{t_n \in T} \Delta t_n W(i,j,t_n) \right) / \sum_{t_n \in T} \Delta t_n . \quad (1.1)$$

Here and below: $\Delta t_n = \Delta t = 10800s$ is the time step, $T$ is the period of averaging, $W(i,j,t_n)$ is the module of wind at the spatial node $(i, j)$ at time $t_n$.

The types of maps are as follows:

a) maps for a winter (January) and summer (July) month, averaged over all years;
b) one-year-averaged maps (for several selected years: 1998, 2001, 2008);
c) a map averaged for the entire period of the wind;
d) a map of trend for one-year-averaged wind speed, estimated for the whole period.

Objectives are to determine the seasonal and annual variability for the wind averaged over different scales; to define a long-term trend of mean wind and its spatial distribution; and, basing on the analysis of the spatial distribution of mean wind, to support the partition of IO into zones.

2. To build and to analyze the maps of mean density for the kinetic-energy-flux of wind (wind power) given by formula

$$< E_A(i,j,T) >= \left( \sum_{t_n \in T} \Delta t_n \rho_a W^3(i,j,t_n)/2 \right) / \sum_{t_n \in T} \Delta t_n \quad , \qquad (1.2)$$

where $\rho_a = (353/T_K)$ $kg/m^3$ is the air density with the account of its dependence on climatic mean temperature $T_K$ (Kelvin) varying with latitude and season. Variation of the argument $T$ corresponds to the four types of averaging mentioned in task 1.

Objectives are to determine the mean density for the kinetic-energy-flux of wind in units of $Wt/m^2$ (further – the wind power) in the waters of IO and its variability in different seasons and years, and to define the climatic trends of wind power and its spatial distribution.

3. To plot the time history of the wind speed averaged over the space

$$W(R,T) = \left( \sum_{t_n \in T} \Delta t_n \sum_{i,j \in R} W(i,j,t_n) \Delta S_{ij} \right) / \sum_{t_n \in T} \Delta t_n / \sum_{i,j \in R} \Delta S_{ij} \quad , \qquad (1.3)$$

and to plot the averaged wind power $E_A(R,T)$

$$E_A(R,T) = \left( \sum_{t_n \in T} \Delta t_n \sum_{i,j \in R} \frac{\rho_a}{2} W^3(i,j,t_n) \Delta S_{ij} \right) / \sum_{t_n \in T} \Delta t_n / \sum_{i,j \in R} \Delta S_{ij} \quad , \qquad (1.4)$$

where $\Delta S_{ij}$ is the area of the grid cell in the area of IO, the left lower corner of which is adjacent to node $i$-th latitude and longitude to the $j$-th one. The argument $R$ implies indexing the spatial domain of integration (a particular point, a part of area, or the whole ocean), the argument $T$ is the averaging period.

It should be constructed:

(a) 11-years' time-history of the "instantaneous" values of wind $W(R,T,t)$ and power $E_A(R,T,t)$ at the fixed points of zones;

(b) time-history for the same quantities, but with their daily averaging (for each zone, and for the entire IO);

(c) daily-mean values of $W(R,T,t)$ and $E_A(R,T,t)$, averaged over the zones and over the entire IO, for the entire 11-year period.

The obtained time series are to be processed by the spectral analysis.

Objectives are to determine the scales of temporal variability of the "instant" and the mean wind and wind power at the fixed points of zones and the IO as a whole.

4. To construct a graph of the time variation of the mean wind $W(R,T,t)$ and wind power $E_A(R,T,t)$, with an annual averaging (for each year of 1998-2008yy) for each zone and the whole IO.

Objective is to determine the 11-year trends of average annual wind and wind power in the zones and the whole IO.

5. To determine the extreme values of wind and their spatial-temporal location (for each zone). To construct maps of the extreme values of wind speed found for each point of the investigated area of IO.

Purpose is to provide information about the real values of maximum winds, including its time and space locations.

6. To plot several kinds of histograms for wind speed:

a) for fixed points of each zone, for the entire period 1998- 2008.

b) for each zone separately with averaging for the whole period;

c) for a dedicated winter (January) and summer (July) months, with averaging for all years and for each zone;

Purpose is to demonstrate the spatial and temporal variability of the distribution function of the module wind.

7. For all variants of the histograms described in task 6, to calculate the four statistical moments (mean, standard deviation, asymmetry and kurtosis) and to estimate the parameters of the model distribution function, parameterized Weibull distribution.

The aim is to demonstrate the extent of correspondence and differences of these statistical characteristics to those known from the literature [6-8].

Note that the specific of our analysis is to study the statistics of the wind field and its energy for the three types of spatial scales: at each point of the ocean (maps); the distribution of characteristics for values averaged over the zones; and the whole-ocean (or integrated) characteristics. This approach can be roughly characterized as enlarging the scale of description of the geophysical field. Farther, in the material consideration, we will adhere to this principle a consistent description of scales.

## 2. Analysis of maps for the fields of wind and its power

### 2.1. Maps of the mean wind

Maps of the mean wind $<W(i,j,T)>$ of all four types, specified in the task 1, are shown in Fig. 1. Their analysis shows the following features of the mean wind fields.

Firstly, all the maps show a clear and stable distribution of wind speed in space. Seen are the following designated areas (referred to as Z#), characterized by the local extremes of average wind speed: the Arabian Sea - Z1, the Bay of Bengal - Z2, equatorial IO - Z3, the Southern-trade-wind part of the IO - Z4, the southern-subtropical part of the IO - Z5, and Southern Indian Ocean - Z6.

Note that if the designated zones Z1 and Z2 are geographically easy to explain (they are separated by the vast Indian subcontinent), as well as the zone of high winds in the Southern Ocean Z6, the presence of zones of weak winds Z3 and Z5, separated by high winds (trade-wind zone Z4), is less predictable. This fact indicates a stable pattern of spatial intermittency of the average wind speed of the IO as a whole, caused by geophysical factors. Such a distribution of mean wind is a rather well-known fact (see [5-8]) that allows us not to dwell on it in more details. It is only important to demonstrate this fact and emphasize it, as it serves as the basis for the further partition of IO into zones with boundaries indicated earlier in section 1.

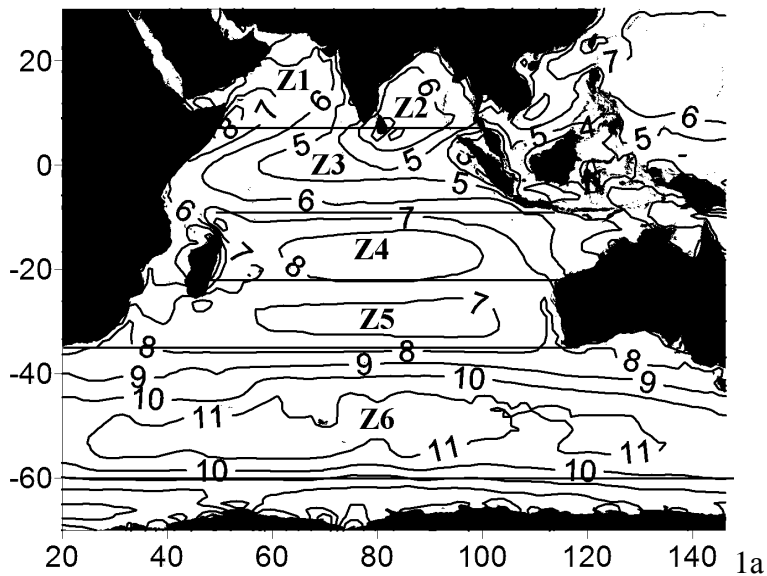

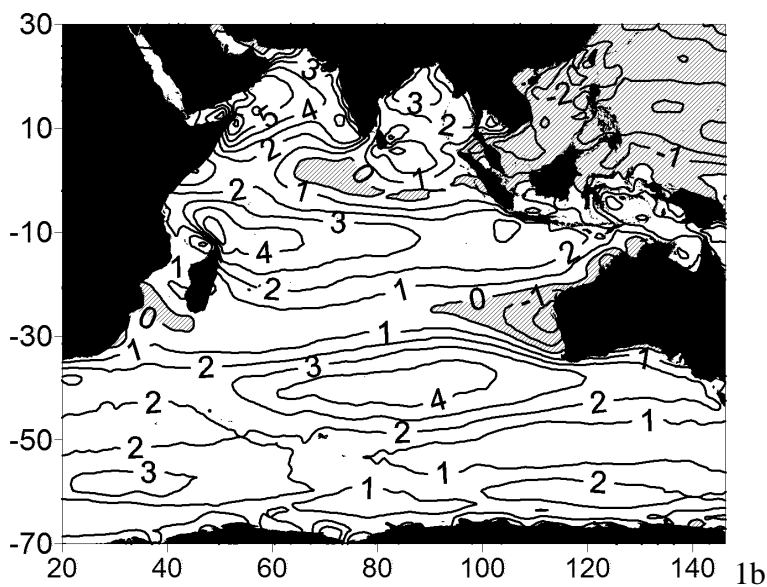

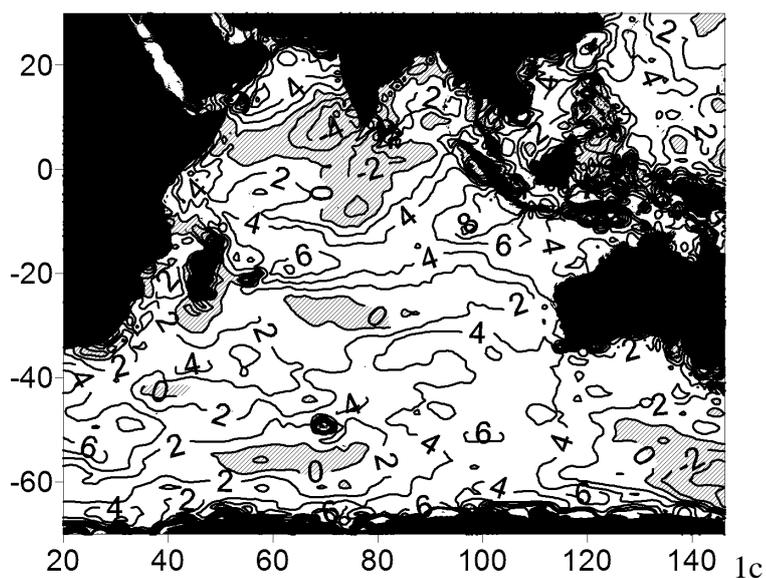

Fig. 1. a) the average wind speed field over the entire period of 1998-2008 years (straight lines indicate the boundaries of zones established by the zoning the Indian Ocean, see Fig. 1-app in appendix), m/s; b) the field of seasonal wind-speed difference (July minus January), averaged over the entire period, m/s; c) trend of the year-averaged wind speed, sm/s/year. Regions of negative values are shaded.

For completeness, we note that the year-averaged wind speed are ranged from minimum values of 4-6 m/s in the equatorial zone Z3, to the maximum 12-14 m/s in the zone Z6 (see Fig. 1-app appendix). These extremes are somewhat smaller than ones averaged for the entire period (Fig. 1a). The certain values of wind, averaged (in addition) over zones and across the ocean, will be given later (Section 4).

Secondly, it is interesting to note a rather strong seasonal variability of the mean field $<W(i,j,T)>$ shown for the example of "summer-winter-variability" (Fig. 1b). Herewith, almost everywhere, except in areas of weak winds Z3 and Z5, the increase in the mean wind in the summer takes place, especially noticeable in the western parts of zones Z1 and Z3. Here, this growth can reach 7-8 m/s, which is associated with the monsoon dynamics taking place in the northern part of the IO. The negative trend of summer winds, which takes place near the west coast of Australia (zone Z5) and reaches -(3-4) m/s, has, apparently, the same nature.

And finally, a few words about the trend of wind, averaged for the entire period (Fig. 1c). The time-trend, obtained by the least squares for each point of the field, is characterized by greater (and even somewhat chaotic) heterogeneity. On average, it is marked the average increase of the wind speed of around 0.3% per year, particularly in areas of high winds (zone Z1, Z4, Z6). Along with this, there are negative trends (in zones Z3, Z5, as well as in the central and eastern parts of zone Z6). However, even the most significant negative trends do not exceed 0.1% per year. All values of trends are in the confidence intervals of these estimates. For a more reliable determination of negative trends (and their explanations), evidently it needs more period of data and a separate discussion.

### 2.2. *Maps of wind power*

The maps of wind power $<E_A(i,j,T)>$ are the obvious derivatives from the wind field. This means that they largely repeat the features of the field $<W(i,j,T)>$, significantly stressing the area of extreme winds. However, in our opinion, the relevance of the field of wind power itself is higher than one for the initial wind field, which is due to a great physical content of the energy concept, compared to the speed one. Just for this reason we put in the foundation of our research study of the wind-kinetic-energy distribution and its variability, as far as this field determines the intensity of the mechanical energy-transfer from the wind to waves, the analysis of which we plan to perform in subsequent papers. In addition, during the assessment of wind power $<E_A(i,j,T)>$, it is also important to take into account the dependence of the air-density on the air-temperature, which was used by us on the basis of data from the report of the European forecasts Centrum [12]. All the foregoing indicates a high self-sufficiency and value of the analysis for the wind-energy field. As we know, calculations and analysis of such a kind of have not appeared in the literature.

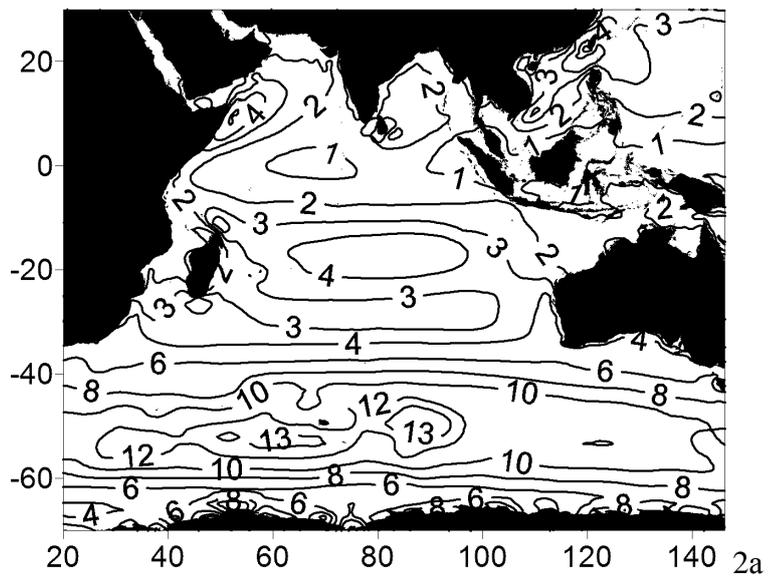
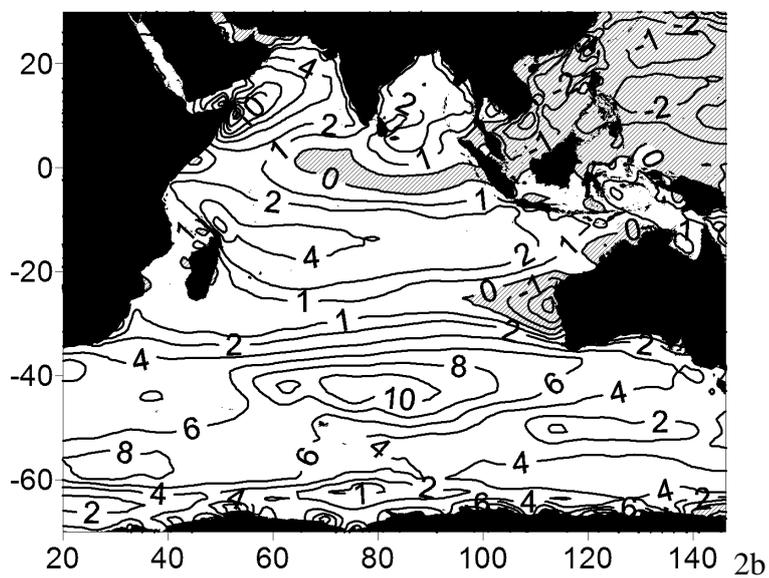
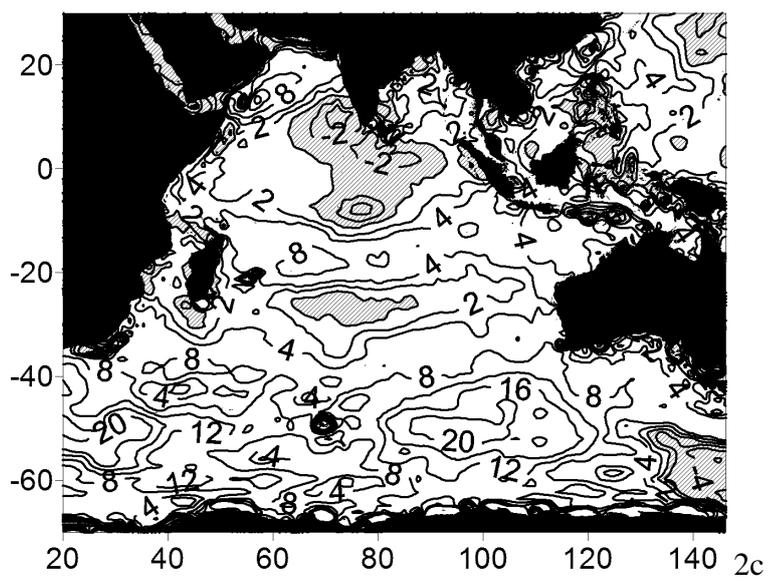

Fig. 2. The same as in Fig. 1,
but for the average wind energy ($10^2$Wt/m$^2$ (a, b), Wt/m$^2$ per year (c)).

The results of calculations of fields of wind power (methodically carried out by analogy with the calculations of wind fields) are shown in Fig. 2. They indicate the following features.

As expected, all the features of spatial and temporal variability of the wind field $<W(i,j,T)>$ becomes more clear in the field of wind power $<E_A(i.j,T)>$, especially in the summer season (see Fig. 2-pp in appendix), and the range of variability in the field of wind power is significantly higher than those for wind speed, reaching a few tens. Note that the wind power reaches several thousands at maximum values (in units of $Wt/m^2$), with the minimum values of order of a few dozen, in coastal areas, and hundreds in the open ocean. These values are fairly consistent with the similar values for the fields $<W(i,j,T)>$ mentioned above.

The said consistency of field values, to a certain extent, takes place for the seasonal, annual, and decadal mean fields and, consequently, for their trends (compare Figs. 1-app and 2-app in appendix). Given the detail in the above description of the fields mean wind $<W(i,j,T)>$, the specified consistency allows less to dwell on the description of details of the fields of wind power.

Taking in to account the pioneering nature of such calculations, to describe the effect of enhancing the dynamic range of spatial variability in the field of wind power, here we give some quantitative estimates of seasonal, interannual, and climate variability. Thus, the range of seasonal variability reaches, at the minimum, values of order $100 Wt/m^2$, and to, at the maximum, $600 Wt/m^2$ (Fig. 2b). Herewith, the inter-annual trend has a range of [-10:50] $Wt/m^2$ per year (Fig. 2c). Moreover, the main increase in energy occurs in the central and middle-east parts of the zone Z6, while a weak negative trend is typical for the equatorial zone Z3 and the far-eastern part of the zone Z6. The reasons for this should be sought in the overall dynamics of atmospheric circulation. It is not excluded that the said is the manifestation of the climate variability, whose study is still far from complete and requires more diverse effort.

At the end of the wind-power field analysis, it should be noted that for purposes of comparison of variability for wind energy and wave energy fields, the fundamental interest have the following aspects:

(a) the spatial distribution of areas of local extremes of the seasonal (annual, long-term) fields for wind and wave energy;

(b) the quantities of local extremes for seasonal and interannual trends of these fields and the variability of their positions;

(c) the time-scales of variability of these fields.

Therefore, in future, after calculating the fields of wave heights, there will probably be a need for a more detailed description of the above aspects of the spatial and temporal variability of fields. However, in this paper, they are not considered.

### 3. Analysis of temporal variations of wind and its power. The scales of their variability

*3.1. Time history of wind speed at fixed points of zones (3-hours step)*

The typical time history of the "instantaneous" wind speed in the central points of zones of IO, coordinates of which are given in Table. 1, are shown in Fig. 3-app in Appendix (step in time is 3 hours). Well seen is a clear change of the wind speed variability towards to the south. If the northern areas show a marked irregularity of the diagrams (i.e. the presence of several scales of variability: a week, season, half year, year), the main scale of variability in the Southern Ocean is the annual one. These visual findings are fully confirmed by the results of spectral analysis of the series (Fig. 3).

Table 1.

The central-point coordinates for the zones in the Indian Ocean

| Zone # | latitude | longitude |
|---|---|---|
| 1 | 14N | 62,5E |
| 2 | 15N | 88,75E |
| 3 | 0N | 80E |
| 4 | 17S | 80E |
| 5 | 29S | 80E |
| 6 | 49S | 80E |

Spectral analysis of the time series was performed using the method of auto-regression analysis based on the Yule-Walker equations, advantages of which are described in detail in [13], and the specifics of which we shall not dwell here. The confidence intervals are of 20% in the logarithmic scale (what is visually seen as a width of spectral curve at the high frequency tail). From the analysis of the frequency spectra population shown in Fig. 4, we can draw the following conclusions.

In zone Z1 (and similar in Z2) the greatest variability takes place for a period of 0.5 year, which is followed by a sufficiently smooth spectrum up to the scales of 1 day, with a weakly isolated peak at the period of 100 days (seasons). On the scales from 100 days to 1 day, the spectrum has a form $S(\omega) \propto \omega^{-1.6\pm0.2}$, which is similar to the Monin-Obukhov law for the decay spectra of isotropic turbulence[14]. At smaller scales there are several sharp peaks at the periods of 1, 0.5, and 0.3 days.

For zone Z3 the general form of the spectrum is retained, but a noticeable peaks take place at the period about 40 days (month variability), and on the scales of 1, 0.5, and 0.3 days. The latter peaks are relatively increased in their intensity, in comparison with zone Z1 (especially at semi-diurnal period).

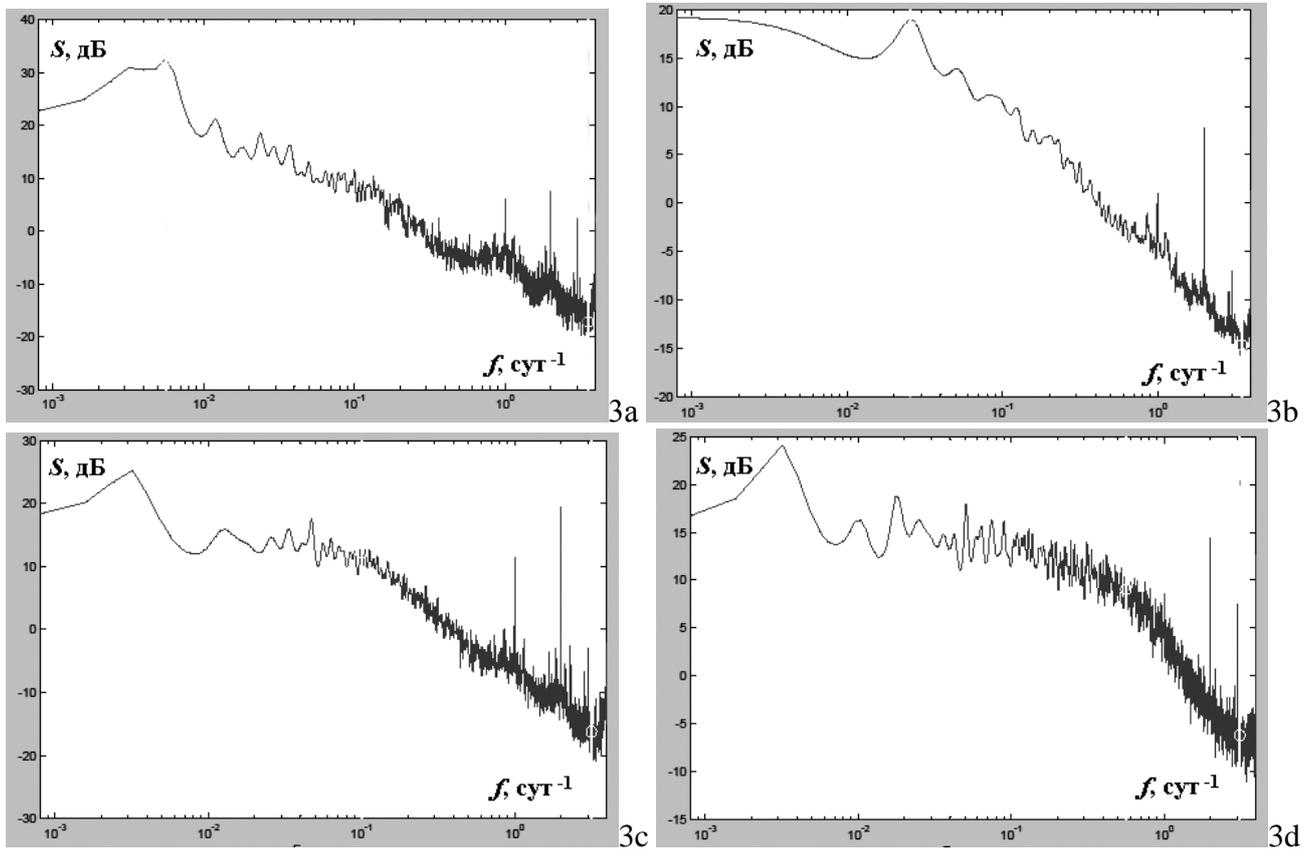

Fig. 3. The spectra of 3-hour series of wind speed for the central points of zones of IO:
a) zone Z1; b) zone Z3; c) zone Z4; d) zone Z6.
The spectral power is given in decibel, the frequency axes is done in inverse days .

For zone Z4 the annual maximum is already well distinguished, which is continued by several small peaks of divisible harmonics: 0.5 and 0.3 year. After that, the spectrum of "white noise" takes place on scales from 100 to 10 days. At more high frequencies, a rather smooth falling spectrum of kind $S(\omega) \propto \omega^{-2.\pm 0.2}$ takes place, having a power-slope somewhat higher than "-5/3". At the tail of the spectrum, there are very sharp peaks on scales of 1, 0.5, and 0.3 days. A similar situation is observed for the spectrum of wind in zone Z6, differing only by an extension of the "white noise" spectrum (with scales ranging from 100 up to 3 days). The slope of the high-frequency part of the spectrum is close to the law "-5 /3".

### 3.2. Time history of the wind power at fixed points of zones (3-hours step)

It is interesting to compare the above results with those but for wind power $<E_A(i.j,T)>$. As the time history of $<E_A(i.j,T)>$ repeats the main features of time evolution for wind speed, being enhanced only with the scale of amplitudes, the main interest has not the time series (not shown) but rather the spectra (Fig. 4).

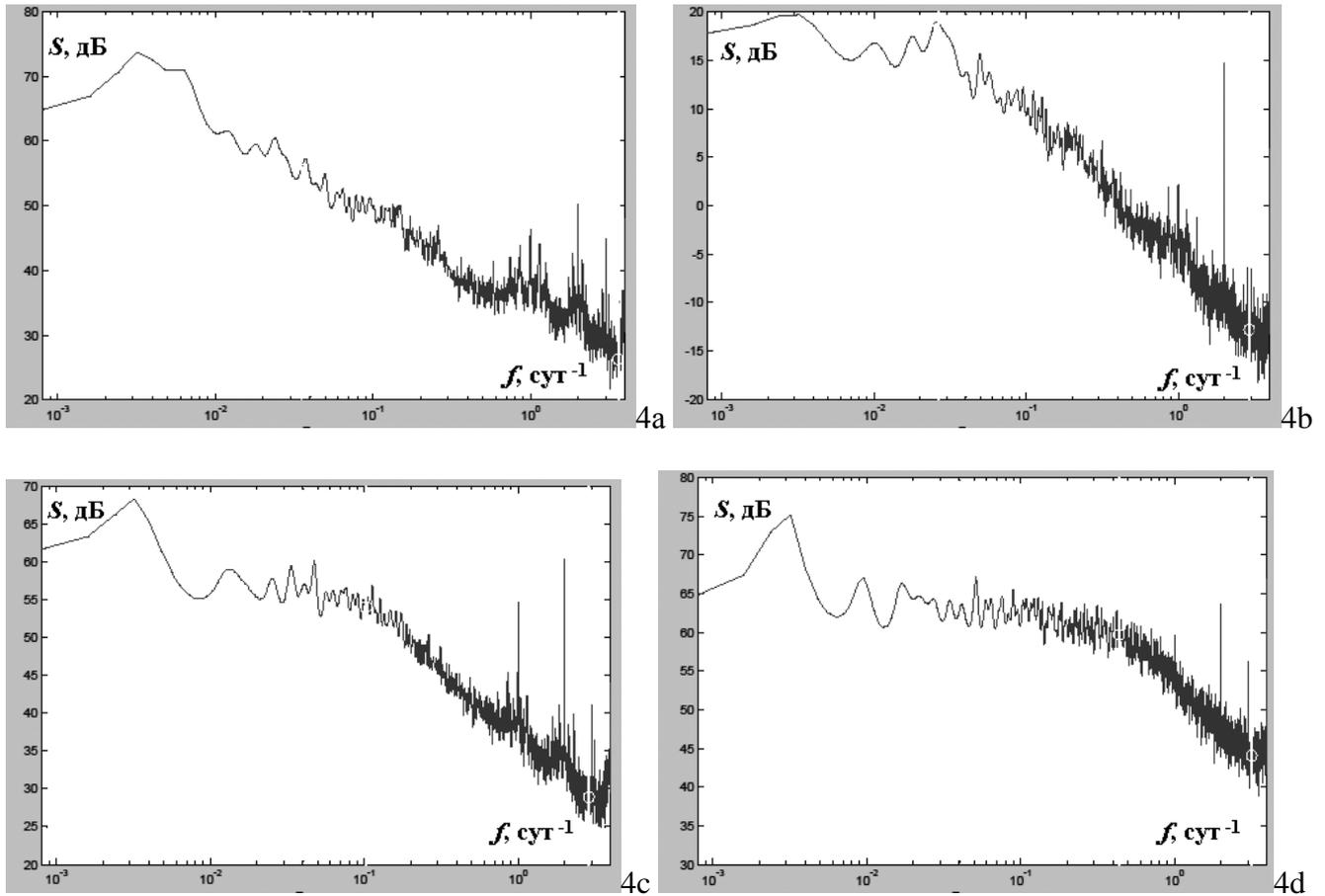

Fig. 4. The same as in Fig. 3, but for the wind power.

So, for zone Z1(and similarly for Z2), the intensity of the annual variability is becoming prevalent, whilst the divisible harmonics are also more or less expressed. Further, from the period of 100 days, the spectrum falls smoothly with the same "classic" law $S(\omega) \propto \omega^{-1.5\pm0.2}$, revealing a well-marked maxima at the scales 1, 0.5, and 0.3 days.

In zone Z3, the spectrum of wind power repeats almost completely the spectrum of wind velocity in this zone.

Spectra in zones Z4-Z6 have the main maximum on scale of 1 year (without harmonic of 0.5 year), followed by the "shelf" (i.e. the spectrum of white noise) on scales from 100 to 10 days (in zone Z4) or less (up to 3 days, in zone Z6). In spectra for all zones, the "shelf" is followed by the turbulent spectrum with the slope of the order of "-2" on high frequency tails. The sharp peaks are visible on the scales 1, 0.5, and 0.3 days (very strong in zone Z4).

Completing the description of variability in the central points of zones, we note that taking into account the similarity of the studied time-history for the wind speed and wind power, for the sake of brevity, in the case of data averaged for a one-day period and over the zone area, we will focus father only on the analysis of temporal variations of the wind power, $<E_A(i.j,T)>$.

### 3.3. Time history of wind power averaged for a day and over the zones

The results of the analysis of this kind are presented in Fig. 5. They imply that as we move to the south, the well expressed semiannual variation, noticeable in the areas of Z1-Z3, becomes gradually less visible. However, the annual variability of wind energy is expressed very strongly in all areas (Fig. 8). This is confirmed by the spectral analysis (Fig. 5).

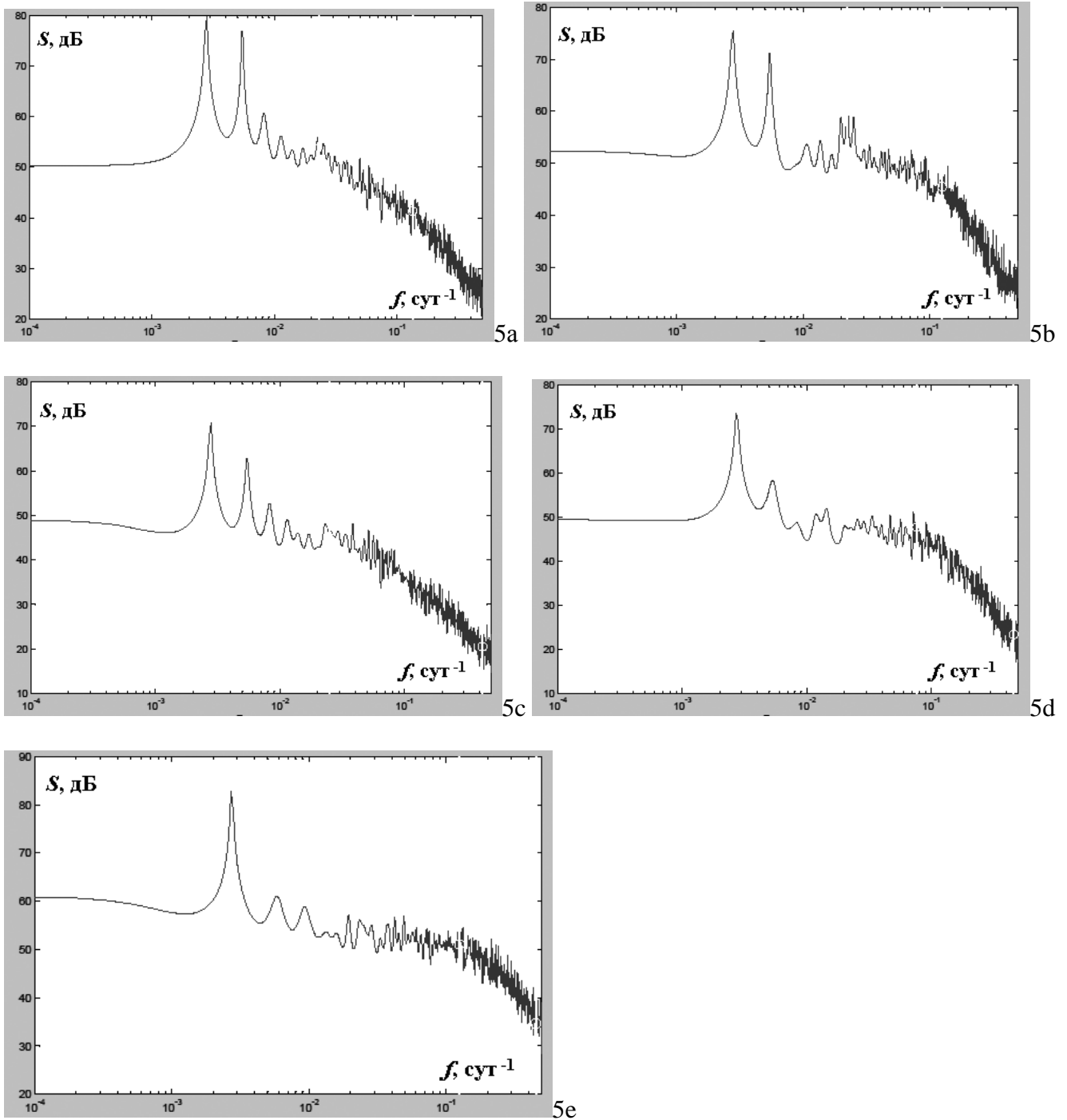

Fig. 5. Spectra for series of wind energy, averaged over one day and over the whole zone:
a) zone Z1; b) zone Z2; c) zone Z3, d) zone Z4; e) zone Z6.

b)

The spectra of temporal variations of wind power in zones Z1-Z3 contain sharp peaks at periods of 1y and 0.5y, which are then replaced by a smooth decrease of intensity in the range of scales from 60 days to 10 days, where the slope of the spectrum is given by ratio $S(\omega) \propto \omega^{-1.6\pm0.3}$. On the scales from 10 to 2-3 days, the slope of the spectrum becomes steeper: $S(\omega) \propto \omega^{-2.5\pm0.5}$.

In the spectra for zones Z2, Z3, a noticeable local maximum at periods of 40 days takes place, which is gradually shifted to period 60, as we move to the south (zone Z4-Z6). Further, on smaller scales, moving from the local maximum to periods of 10 days, the white noise spectrum is realized. For shorter periods, there is a constant slope spectrum of the kind $S(\omega) \propto \omega^{-2.5\pm0.5}$.

In conclusion, it is interesting to consider the time history (Fig. 4-app in appendix) and the spectrum of the wind and wind power, averaged for a day and over whole IO (Fig. 6).

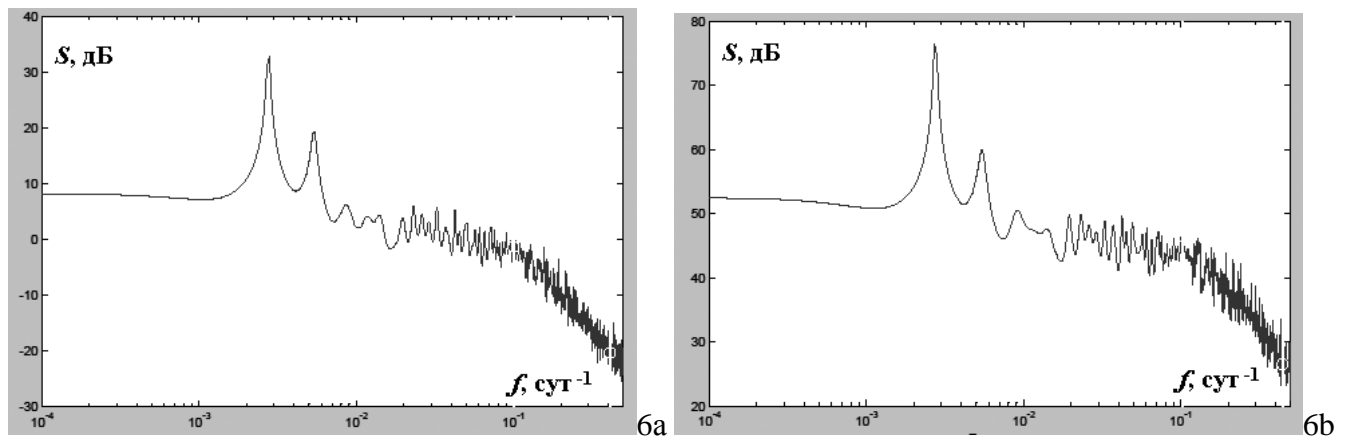

Fig. 6. Spectra series of wind speed (a) and energy (b),
averaged over a day and across the whole IO.

First, the wind power averaged over whole IO shows a very strong variability (taking into account the size of the ocean). This variability on scales of several days can change up to 1.5 times (upper panels of Fig. 4-app). Such a strong variability of wind power over the whole ocean is seen as a very unexpected result.

Secondly, the spectrum of this series (Fig. 6) has only two very clearly marked periods: 1 year and 0.5 year. Then, there is a long "shelf" from the period of 100 days to 10 days, which is replaced by a rather sharp decrease of the power-law form of the kind $S(\omega) \propto \omega^{-2.5\pm0.3}$.

Third, at the tail of the spectrum there are visible peaks corresponding to the scales of 8-7days and 5-3days. Despite the fact that these peaks are "drowning" in the statistical noise of the spectral estimates, we believe that just this small-scale variability corresponds the above-noted irregularity of the total wind power of the ocean (Fig. 4-app).

*3.4. Conclusions*

Naturally, the detailed analysis of the spectra and their interpretation require a separate and more comprehensive study and discussion. Here, based on analysis of these figures, we restrict ourselves to formulation of the following conclusions.

3.4.1. Registered are the following scales of variability (both for wind speed and its power): 1 year, 0.5 year, 100 days, 60 days, 40 days, 1day, 0.5 days, and 0.3 days. In the spectrum for wind-power averaged over whole Indian Ocean, the scales 1 year, 0.5year have the main importance.

3.4.2. In the northern areas of IO, the wind speed has the main variability on the scale 0.5 year, whilst in the southern IO does 1 year.

3.4.3. For the average wind power, between the mentioned scales of maximum variability and the small scales of turbulent nature, the spectrum has a domain of "white noise" behavior. This domain begins from the scales of 100 days and ends at scales of 10-3 days. This result indicates the absence of a correlation between motions at the scales of the specified range. This feature of the wind field variability is established for the first time.

3.4.4. At scales smaller than the scales of the white noise domain, there is a power-law decay of spectrum. For the spectrum of wind speed it is close to the law "-5 / 3"; for the wind-power spectrum, the decay rate of the spectrum tail is slightly larger and lies in the range of -2 to -2.5 (with an accuracy of estimation of the order of 10-15%).

3.4.5. At the tails of spectra for the fields averaged over whole IO, there are remarkable peaks corresponding to scales of 7-8 days and 3-5 days. This middle-scale variability reflects the fact of strong time-irregularity of the series for wind and wind-power fields averaged over whole IO (Fig. 4-app).

## 4. Analysis of the integral and point features

*4.1. Time history for the annual-average values*

On two panels of Fig.7 there are shows the time history of the wind speed (upper) and the wind power (lower panel), annually averaged in time and averaged over zones and entire IO. It can be seen a quite expected distribution of the average values among zones. It is interesting to note that estimates of the annual-average wind power almost correspond to the cube of the annual-average wind speed for each zone. It means that after averaging, the ration (1.4) takes the kind

$$E_A(R,T,t) \cong (0.8-0.9)W^3(R,T,t) \qquad (4.1)$$

This empirical fact can be used to obtain quick, so called "expert" estimates of the average wind power, without recourse to more cumbersome calculations.

Confining ourselves to the analysis of the integral value of wind energy for whole IO, we should note two points: a) presence of a barely discernible trend for the average wind speed: it has the growth of the order of 0.25% per year; (b) a quite noticeable trend in the mean wind power over IO: an increase of 1% per year. According to values of $R^2$, the trend for wind is not reliable, whilst the trend for wind power is reliable. It means that the last result is of major interest in the light of climate variability of wind power in the waters of IO.

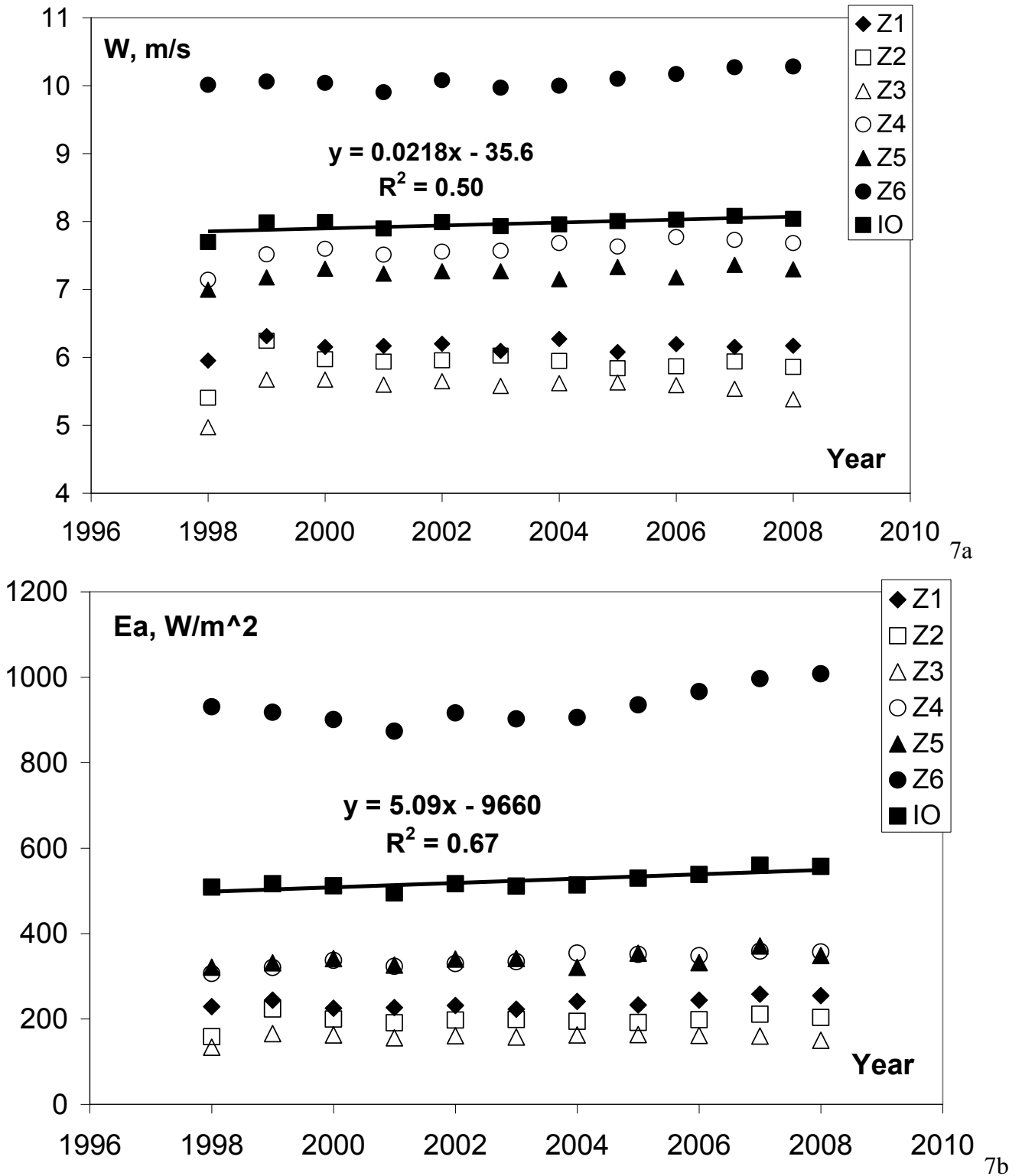

Fig. 7. Time variation of speed (a) and energy (b) of wind, averaged for a year and over the zones (including the entire area of the IO as the average for all zones).
Shown are the r.m.s. trend lines for the entire area of IO and their formulas.
Criterion $R^2$ is present (R is the ration of the trend value to the standard deviation of series).

*4.2. Location of extremes of wind speed*

Magnitudes of extremes of wind speed and coordinate of their spatial and temporal distribution are given in Tab. 2. It needs especially to note that these space-time coordinates are of interest for comparison with similar coordinates of the maxima for wave heights. Such a comparison is of fundamental importance both for demonstrations and for assessment of the space-time separation for the extremes of wind and wave fields.

Analysis of the table indicates on the following.

First, the largest magnitudes of wind, exceeding values of 32-35m/s, are realized in the most southern zone of IO. Moreover, these maxima take place mainly during the summer period (on the calendar of the northern hemisphere). Note here that those absolute maximum which are located outside the $60^0$S-latitude are not shown, as we have adopted $60^0$S as the background ice edge in IO.

Secondly, the absolute maximum of 42.1 m/s takes place in the "roaring fortieth" (coordinates: 47S, 90E) on the date 21.06.1998. Just these coordinates are of primary interest for comparison with the analogous coordinates of the absolute maximum wave heights, which will be estimated in future.

Thirdly, in zones Z1-Z4 the 10-year maxima of wind are of the order of 23-25 m/s. This is fairly large values what itself is of our interest as a characteristic of wind regime in these areas. However, locations of these extremes are rather randomly distributed in space and time, which allows us here to restrict ourselves by a statement of this fact only.

Complete picture (chart) of the wind maxima is shown in Fig. 5-app as "synthetic" maps, points of which are not tied to a single time.

*4.3. Histogram of wind speed (in the zones and the entire IO)*

The most complete probabilistic information about the field of wind speed is given by the histograms, the study of which is devoted in a lot of papers (see references in [5-8]). The histograms are the numerical representation of the probability density functions (PDF). For many geophysical fields, PDF are usually parameterized by one or another kind of Weibull distribution, which is given by the formula[7]

$$P_{n,a}(W) = \frac{n}{a}\left(\frac{W}{a}\right)^{n-1} \exp\left[-\left(\frac{W}{a}\right)^n\right], \qquad (4.2)$$

where *n* is the dimensionless "power" of the distribution, and *a* is the "scale" parameter. Parameters of the distribution, *n* and *a*, vary greatly in space and time, and, naturally, depend on the size of the space-time averaging of the wind-field studied. Such variability of the probabilistic structure of a wind-field reduces significantly the importance of parameterization (4.2). Nevertheless, the parameterization of histograms in the form (4.2) is of particular interest for construction of stochastic models for real wind fields, resulting from the principal dynamic equations, as a part of understating the causes of their variability [8].

Table 2.

Wind speed maxima in the zones for the Indian for the whole period

| Year\ Zone | 1 | 2 | 3 | 4 | 5 | 6 |
|---|---|---|---|---|---|---|
| 1998 | **20.3** <br> 09.0\| 68.75 <br> 13/12 | **24.5** <br> 21.0\| 88.75 <br> 10/07 | **23.9** <br> 0.0\| 47.50 <br> 21/05 | **31.3** <br> -16.0\| 61.25 <br> 09/02 | **32.9** <br> -30.0\| 70.0 <br> 13/02 | **42.1** <br> -47.0\| 90.0 <br> 21/06 |
| 1999 | **19.5** <br> 10.0\| 52.50 <br> 07/07 | **20.0** <br> 19.0\| 88.75 <br> 28/10 | **21.0** <br> -8.0\| 75.00 <br> 20/10 | **22.2** <br> -12.0\| 73.75 <br> 08/10 | **23.2** <br> -35.0\| 52.50 <br> 17/05 | **29.6** <br> -48.0\| 83.75 <br> 23/07 |
| 2000 | **21.1** <br> 12.0\| 56.25 <br> 25/07 | **21.6** <br> 9.0\| 87.50 <br> 24/12 | **23.9** <br> -9.0\| 82.50 <br> 05/01 | **31.7** <br> -13.0\| 50.00 <br> 01/03 | **26.5** <br> -33.0\| 31.25 <br> 03/09 | **30.7** <br> -43.0\| 21.25 <br> 23/04 |
| 2001 | **21.4** <br> 19.0\| 62.50 <br> 05/03 | **18.3** <br> 17.0\| 83.75 <br> 13/06 | **19.6** <br> -7.0\|101.25 <br> 09/02 | **25.3** <br> -15.0\| 71.25 <br> 10/01 | **32.3** <br> -33.0\|125.00 <br> 01/02 | **32.8** <br> -37.0\| 47.50 <br> 12/03 |
| 2002 | **18.2** <br> 10.0\| 52.50 <br> 12/06 | **21.0** <br> 17.0\| 93.75 <br> 18/05 | **16.7** <br> 2.0\| 95.00 <br> 08/05 | **21.2** <br> -12.0\|112.50 <br> 12/04 | **24.4** <br> -32.0\| 56.25 <br> 13/03 | **32.4** <br> -56.0\| 52.50 <br> 30/07 |
| 2003 | **19.8** <br> 21.0\| 68.75 <br> 27/07 | **21.9** <br> 16.0\| 93.75 <br> 23/07 | **17.2** <br> 6.0\| 50.00 <br> 07/07 | **21.4** <br> -22.0\| 66.25 <br> 13/03 | **25.9** <br> -35.0\| 95.0 <br> 28/05 | **34.7** <br> -53.0\|116.25 <br> 13/04 |
| 2004 | **22.2** <br> 16.0\| 66.25 <br> 14/06 | **19.9** <br> 16.0\| 85.00 <br> 13/06 | **23.0** <br> -9.0\| 73.75 <br> 24/11 | **25.8** <br> -12.0\| 72.50 <br> 25/11 | **26.4** <br> -32.0\| 70.0 <br> 29/05 | **33.3** <br> -57.0\| 55.0 <br> 21/08 |
| 2005 | **20.2** <br> 20.0\| 68.75 <br> 16/09 | **19.8** <br> 15.0\| 91.25 <br> 06/07 | **16.7** <br> -5.0\| 80.00 <br> 28/01 | **19.7** <br> -16.0\| 90.00 <br> 24/11 | **25.9** <br> -35.0\| 67.50 <br> 23/08 | **34.6** <br> -46.0\| 58.75 <br> 19/08 |
| 2006 | **21.6** <br> 14.0\| 73.75 <br> 29/05 | **23.5** <br> 16.0\| 88.75 <br> 02/07 | **20.6** <br> -9.0\| 51.25 <br> 22/12 | **22.9** <br> -19.0\|113.75 <br> 21/01 | **25.9** <br> -35.0\| 62.50 <br> 03/05 | **34.9** <br> -54.0\| 46.25 <br> 30/12 |
| 2007 | **23.7** <br> 20.0\| 71.25 <br> 02/07 | **26.6** <br> 15.0\| 86.25 <br> 28/06 | **20.0** <br> 6.0\| 51.25 <br> 18/06 | **28.3** <br> -22.0\| 53.75 <br> 28/02 | **29.8** <br> -30.0\| 51.25 <br> 01/03 | **34.3** <br> -45.0\| 87.50 <br> 27/07 |
| 2008 | **21.3** <br> 20.0\| 71.25 <br> 12/08 | **21.4** <br> 18.0\| 91.25 <br> 15/06 | **17.1** <br> 6.0\| 50.00 <br> 05/08 | **24.4** <br> -19.0\| 51.25 <br> 17/02 | **27.5** <br> -32.0\| 37.50 <br> 21/09 | **34.3** <br> -44.0\|110.0 <br> 27/07 |

Note. The form of data presentation: the upper value is the maximal wind $W_{max}$ (bold, in units m/s), next line is the latitude|longitude of the maximum, the lowest line is the time (day/month) of the maximum origin. The sign "minus" means the southern hemisphere.

The histogram gives a basis for calculation of statistical moments formally given by the formula

$$M_i(W) \equiv <W^i> = \int_0^\infty W^i P(W) dW = a^i \Gamma(1+i/n) , \quad (4.3)$$

where $i$ is the number of the moment, and $\Gamma(...)$ is the gamma function. Just these statistical moments are of particular practical interest[7].

So, the first moment, $<W>$, has a meaning of the mean wind value, $M$. It is given by the expression

$$M = a\Gamma(1+n^{-1}). \quad (4.4)$$

The second moment $<W^2>$ gives the integrated information about the scatter of data, more definitely, it does the standard deviation (and dispersion) of the distribution, $D$:

$$D(W) = \left[\int_0^\infty (W-<W>)^2 P(W) dW\right]^{1/2} = a\left[\Gamma(1+2n^{-1}) - \Gamma^2(1+n^{-1})\right]^{1/2}, \quad (4.5)$$

The third moment $<W^3>$ permits to calculate the skewness (or asymmetry) of the distribution, $A$, given by the formula

$$A = \int_0^\infty (W-<W>)^3 P(W) dW / D^3(W), \quad (4.6)$$

and the fourth moment does the kurtosis (or excess) of distribution, $E$, given by

$$E = \left[\int_0^\infty (W-<W>)^4 P(W) dW / D^4(W)\right] - 3 . \quad (4.7)$$

Following [7], we note that the values of mean $M$ and standard deviation $D$ of the studied series give the principal parameters of the distribution (4.1) (with an accuracy of about 5%):

$$n \cong [M/D]^{1.086} \quad \text{и} \quad a = M/\Gamma(1+n^{-1}). \quad (4.8)$$

That allows us to identify parameters for PDF of the form (4.2) from the real histogram of the variable under investigation. In turn, the values of $A$ and $E$, characterizing the deviation from the Gaussian distribution function, can describe some features of the PDF under consideration.

Thus, a positive value of asymmetry $A$ means that the mean value $M$ is larger than the most likely value of variable $W$ under consideration (and vice versa); and variable $W$ has a positive (negative) kurtosis, if its PDF is more sharply peaked (broadly peaked) and has longer (shorter) tails than the Gaussian PDF with the same mean $M$ and standard deviation $D$. Positive value of $E$ leads to a "prolongation" of the PDF tail, i.e. it corresponds to an increase of probability for the extreme high values of variable $W$ (and vice versa). Both the skewness and kurtosis are zero for the Gaussian distribution.

Given such a high degree of scrutiny of the wind-field histograms, their analysis is of little value for our purposes. However, for the sake of completeness, we present here a series of their graphical representations (Fig. 8) and a table of distribution for parameters *n*, *a* and moments *M*, *D*, *A* and *E* as the fixed points of the zones, and for the ocean as a whole (Table 3).

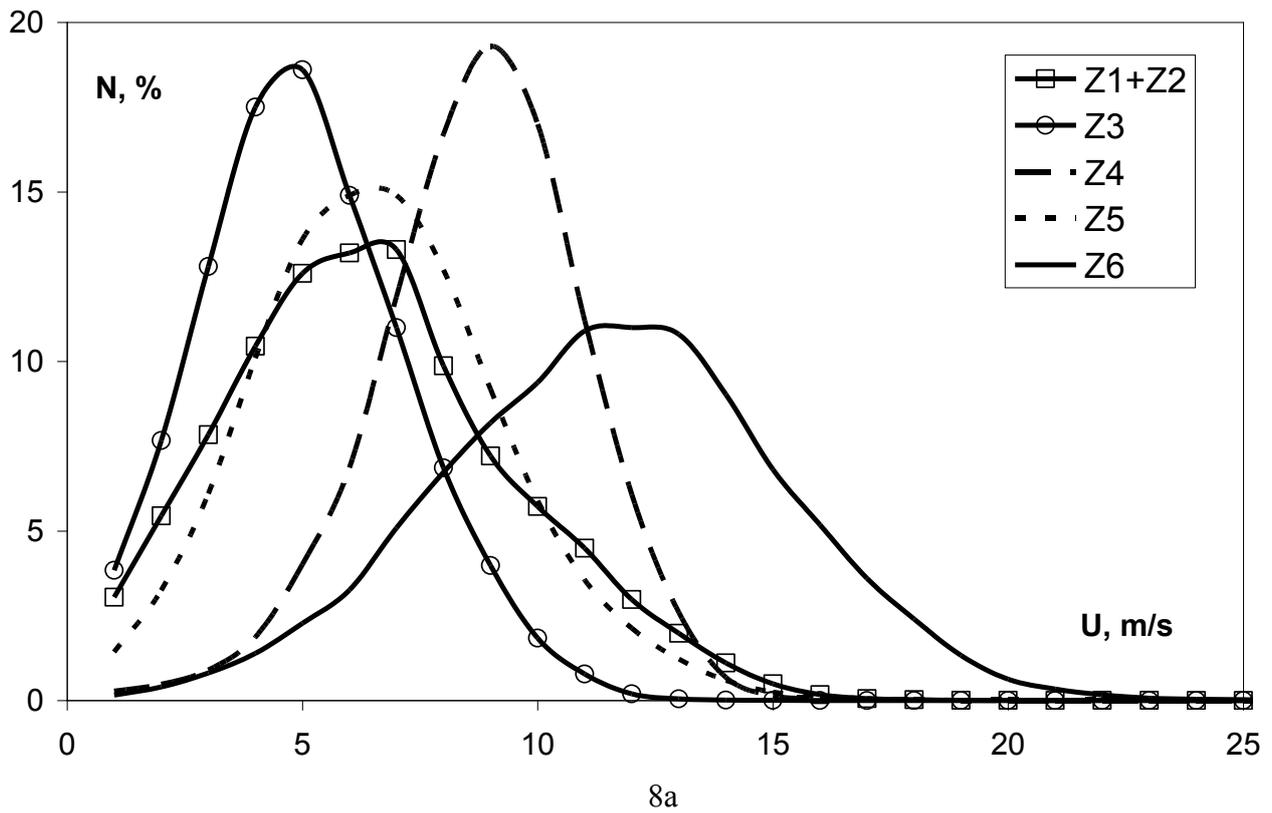

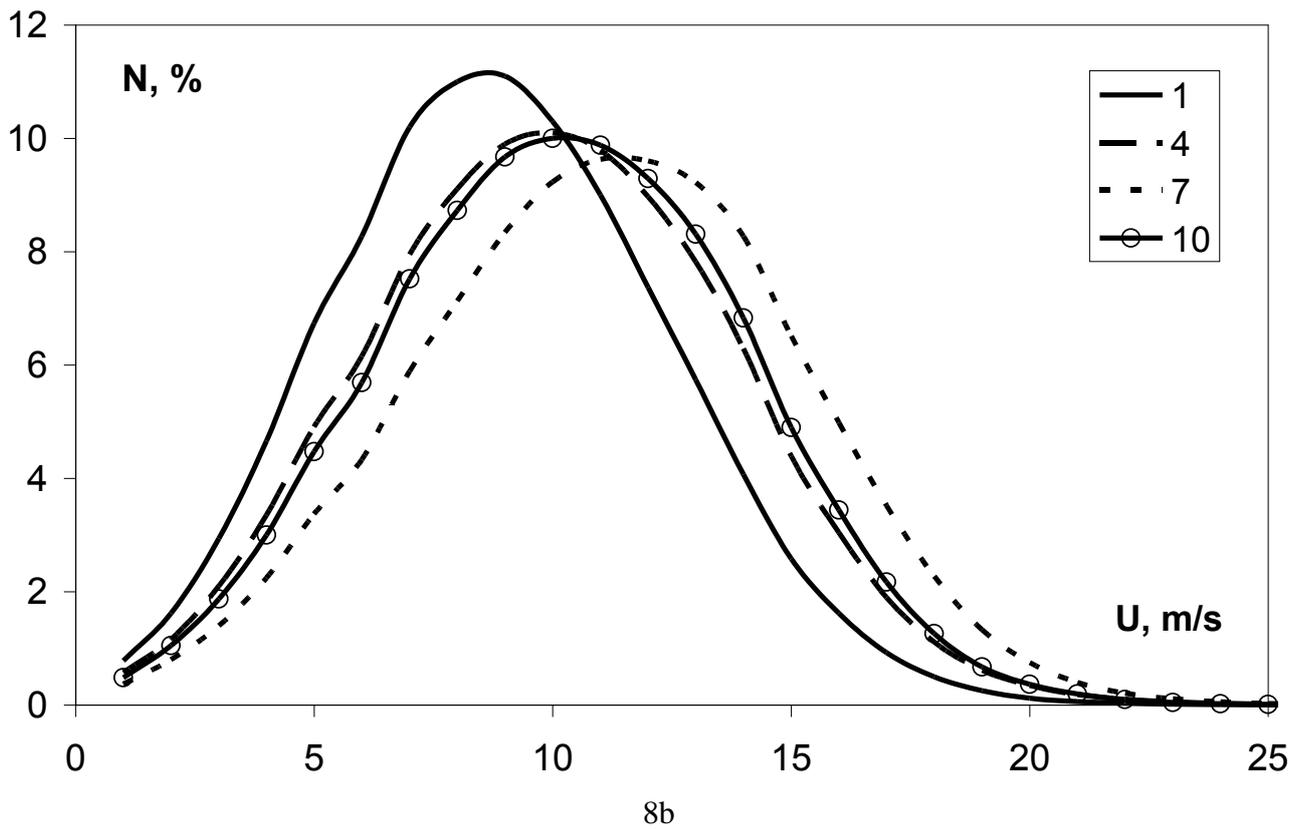

Fig. 8. Histograms of the modulus of wind over the entire period for the central points of the zones (a) and allocated to four months in the entire zone Z6 (b), for clarity presented in the form of lines. Number denotes the month.

Table 3.

Statistical characteristics of the wind field
for central points of the zones, whole zones(seasons and whole period),
and for the whole Indian

| Type of averaging | M, m/s | D, m/s | A | E | *n* | *a*, m/s |
|---|---|---|---|---|---|---|
| $Z_1$ | 6.8 | 3.1 | 0.41 | -0.34 | 2.34 | 7.7 |
| $Z_2$ | 6.3 | 2.9 | 0.47 | 0.22 | 2.27 | 7.1 |
| $Z_3$ | 5.1 | 2.2 | 0.36 | -0.11 | 2.52 | 5.7 |
| $Z_4$ | 8.7 | 2.2 | -0.31 | 0.41 | 4.45 | 9.5 |
| $Z_5$ | 6.7 | 2.7 | 0.43 | 0.48 | 2.7 | 7.4 |
| $Z_6$ | 11.6 | 3.6 | -0.006 | -0.026 | 3.5 | 12.6 |
| Z1-w|s | 5.9 | 9.4 | 2.1 | 2.9 | -0.052| -0.15 | -0.14 | -0.13 | 3.09 | 3.62 | 6.6 | 10.3 |
| Z2-w|s | 5.6 | 7.8 | 1.9 | 2.5 | -0.10 | 0.036 | -0.30 | 0.14 | 3.15 | 3.40 | 6.3 | 8.5 |
| Z3-w|s | 5.3 | 6.8 | 2.3 | 2.4 | 0.24 |-0.16 | -0.15 | -0.33 | 2.49 | 3.05 | 5.9 | 7.6 |
| Z4-w|s | 6.5 | 8.9 | 2.3 | 2.3 | 0.08 |-0.41 | 0.17 | 0.22 | 3.04 | 4.36 | 7.2 | 9.7 |
| Z5-w|s | 7.0 | 7.8 | 2.6 | 3.1 | 0.14 | 0.36 | 0.05 | 0.23 | 2.96 | 2.74 | 7.7 | 8.7 |
| Z6-w|s | 9.0 |11.2 | 3.5 | 3.9 | 0.23 | 0.02 | -0.11 | -0.18 | 2.78 | 3.09 | 10.0|12.4 |
| Z1-tot | 6.2 | 2.9 | 0.58 | 0.14 | 2.28 | 7.0 |
| Z2-tot | 5.9 | 2.6 | 0.36 | -0.061 | 2.43 | 6.7 |
| Z3-tot | 5.6 | 2.4 | 0.21 | -0.35 | 2.51 | 6.2 |
| Z4-tot | 7.6 | 2.5 | -0.17 | -0.047 | 3.39 | 8.3 |
| Z5-tot | 7.3 | 2.8 | 0.35 | 0.33 | 2.76 | 8.1 |
| Z6-tot | 10.1 | 3.8 | 0.17 | -0.17 | 2.85 | 11.2 |
| IO-tot | 7.1 | 3.3 | 0.63 | 0.66 | 2.33 | 8.0 |

Note. Type of averaging: $Z_i$ means using histogram at the central point of the zone i;
Zi-w|s means using histogram for the whole zone Zi in the selected month (w :January, s: July) for the whole period; Zi-tot means using histogram of zone $Z_i$ for the whole period; IO-tot means using histogram for the whole Indian and for the whole period.

Without dwelling on the obvious trends of the average values through zones, consider here the features of skewness *A* and kurtosis *E* volatility with the aim of their comparison with the known values for the Indian Ocean from [7]. It is important to note that in our database, the significant (i.e. greater than the "threshold" value of 0.05) negative values of skewness are observed mainly in zone Z4. In the winter time, this effect takes place also in zones Z1, Z2 (along with zone Z4). Herewith, the quantities of the Weibull distribution parameter *n* have a marked increase.

Negative values of *E* take place in areas Z1, Z3 and Z6, showing, however, considerable variability depending on the scale of averaging, and constraining their binding to the certain zones. Nevertheless, in general, our results significantly clarify and elaborate the earlier results of [7], obtained at much greater spatial and temporal averaging.

In this connection one should pay attention to the result for all IO (the last row in the table. 3). From this it obviously follows the complete unacceptability of the statistical study for the wind field (as well as any other geophysical field) on the scale of the whole ocean, since such an approach does not reflect regional specificity in this field.

### 5. General findings and conclusion

The most general conclusions from the results of this study are as follows.

First, the wind field in the waters of the Indian Ocean is rather inhomogeneous. Therefore, the task of studying the statistical properties of the wind field (as well as other geophysical fields) in this area requires its zoning (Section 2). Estimations of the ocean-averaged values are acceptable only for the study of their variability on the long-term (climate) scale.

Secondly, the scales of variability for "instantaneous" values of the wind in the fixed points and for the series, produced by their spatial and temporal averaging, are rather different. Nevertheless, for certain zones they are fairly stable to variations of scale averaging. The said is applied both to the presence of main scales of variability for a wind field in the zones: 1year, 0.5 year, high divisible harmonics; and to the lack of correlation between the movements on the scales of periods from 100 to 10 days (Section 3). The latter is pioneering result which is especially characteristic to the southern zone of IO and to the series of data averaged over whole IO, what was definitely established by the spectral analysis of proper series.

Third, the study of features for the wind-field kinetic power is of an independent and justified interest. In particular, in this study we estimated its seasonal, interannual and climate variability both for zones and for the whole Indian Ocean. It is shown that for the time interval 1998-2008yy, the total wind power has an average positive trend of about 1% per year; and this result is statistically reliable (Section 4), according to our estimation. This result correlates well with the published information about the impact of global warming on the intensity of atmospheric circulation (see references in [3,5]). Of cause, it needs a further studying in this topic with using data for longer periods.

In conclusion we note that the proposed here method of data processing will be used for a similar analysis of the wave-field in IO, calculated for a given wind field. Such an analysis would identify the relevant statistical characteristics and their spatial and temporal distribution. Comparison of information obtained for different geophysical fields will reveal the patterns of their relationships at different scales, whose nature is little understood yet.

The authors thank academician Golitsyn G.S. and chief of the project Prof. Ginzburg A.S. for many discussions of issues arising in the course of the project. We grateful to Mokhov I.I. for his constant interest to the topic, and Kostrykin S.V. for his assistance in unpacking the wind data. This work was supported by RFBR, grant № 10-05-92662-IND_a.

**Appendix**

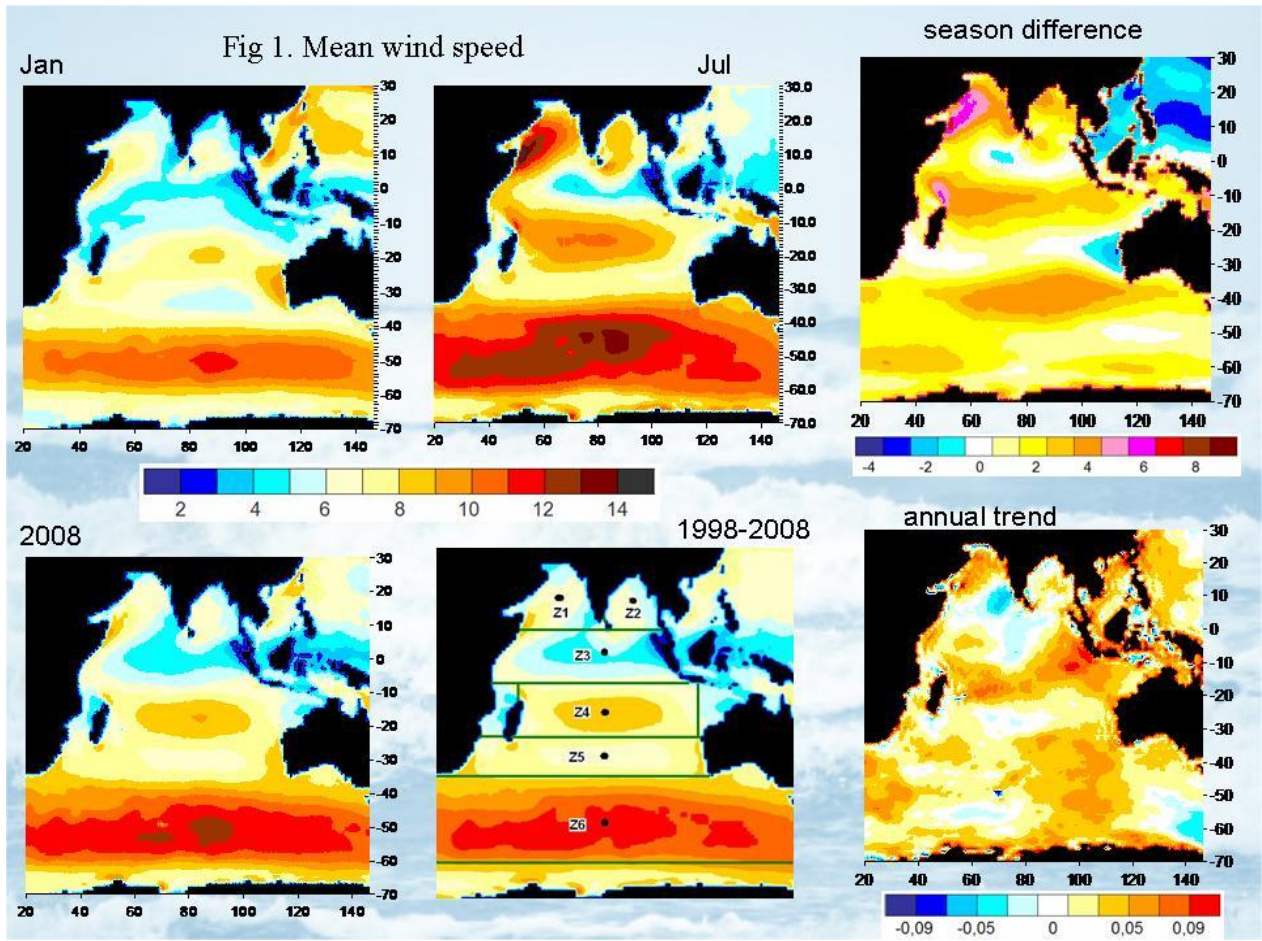

Fig. 1-app. The maps of wind speed averaged on different scales.
Left-upper panels : January and July for all the period; Left-lower: averaging for 2008year; right-upper panel is the season difference (see text); right-lower panel: annual r.m.s. trend.
The first and second scales are given in m/sec, the third is done in percent per year.

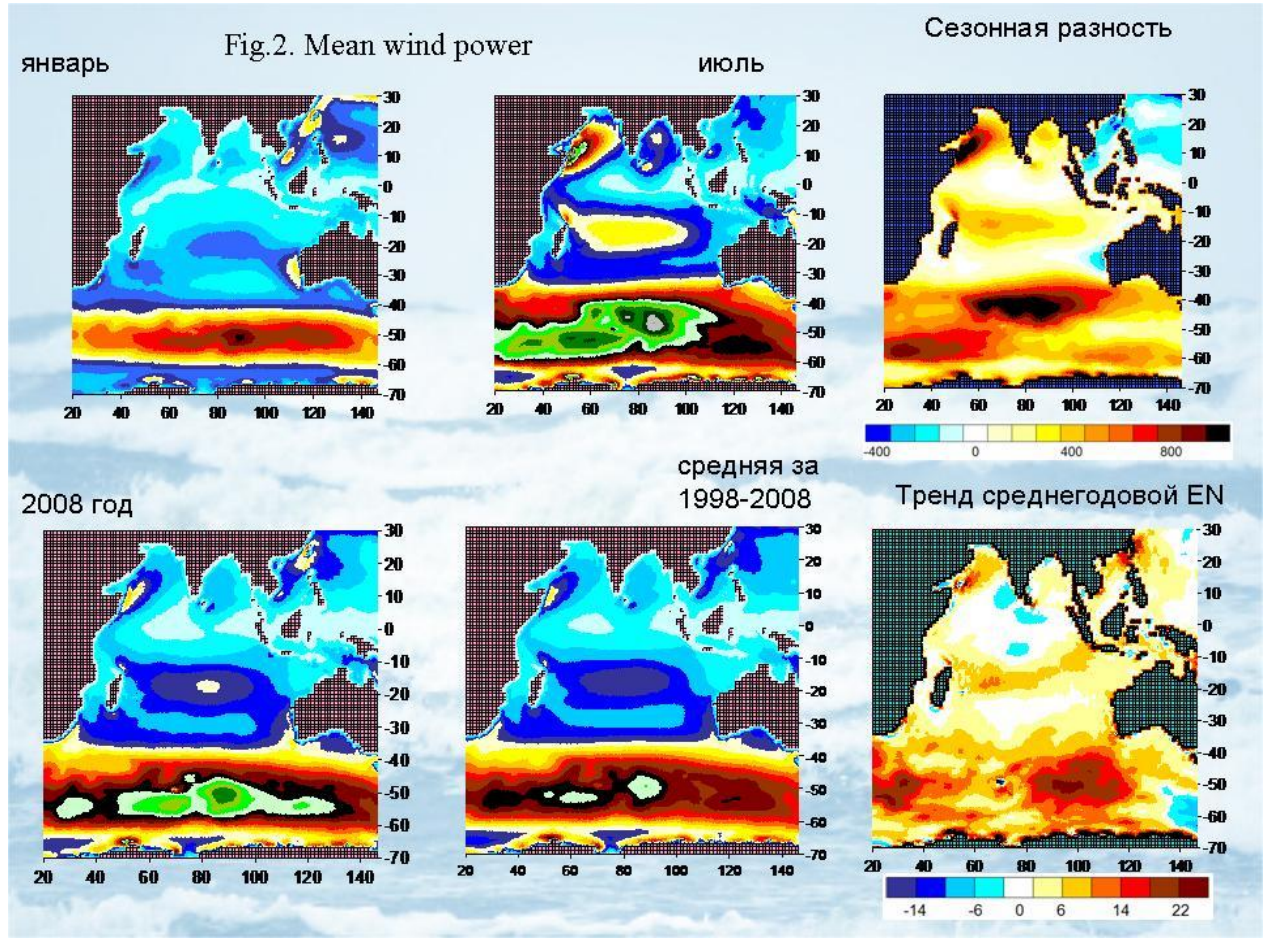

Fig. 2-app. The same as in Fig. 1 but for the wave power.
The upper scale is given in W/m$^2$, the lower scale is done in W/m$^2$ per year.

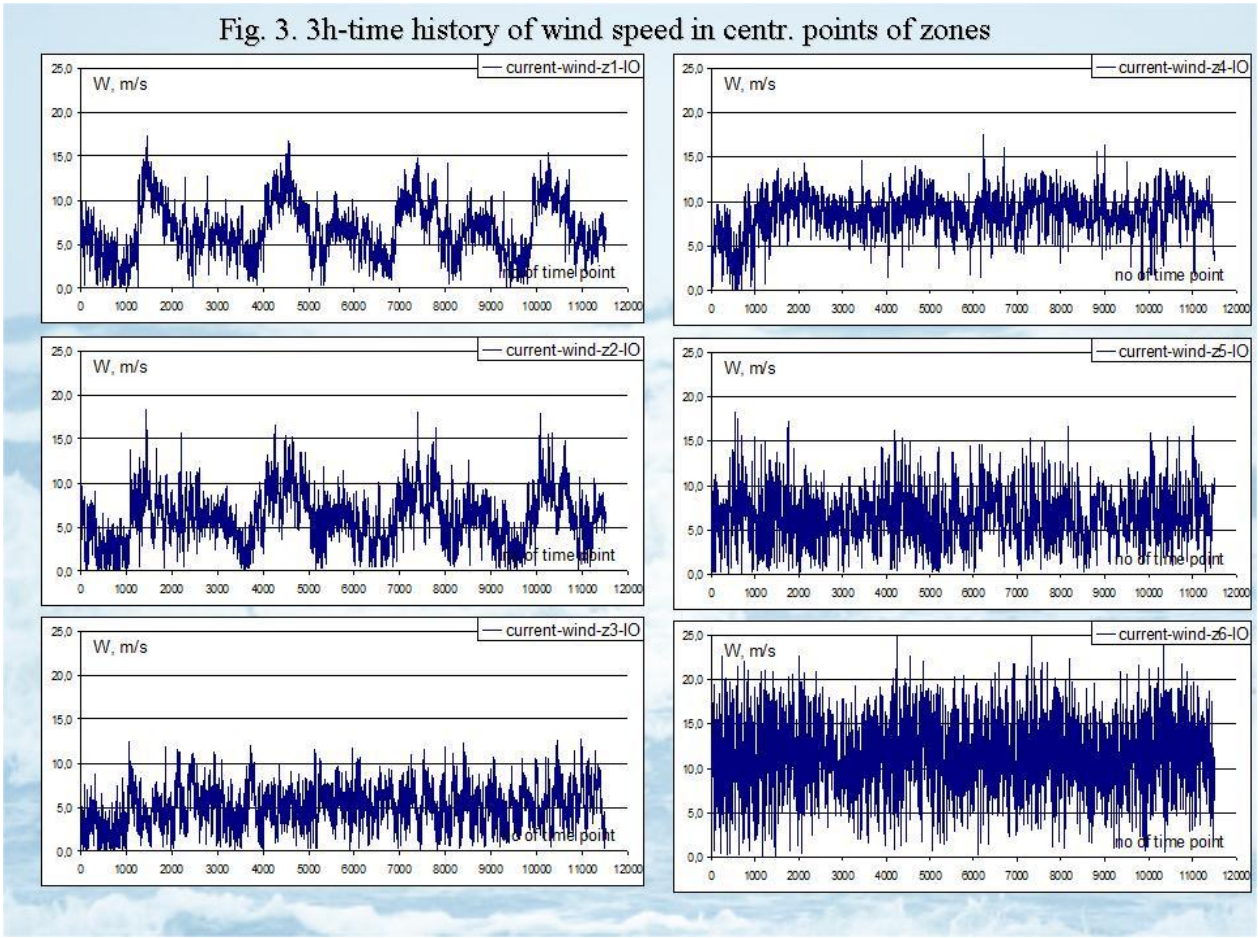

Fig 3-app. The time history of wind speed at the central points of zones (3h-step).

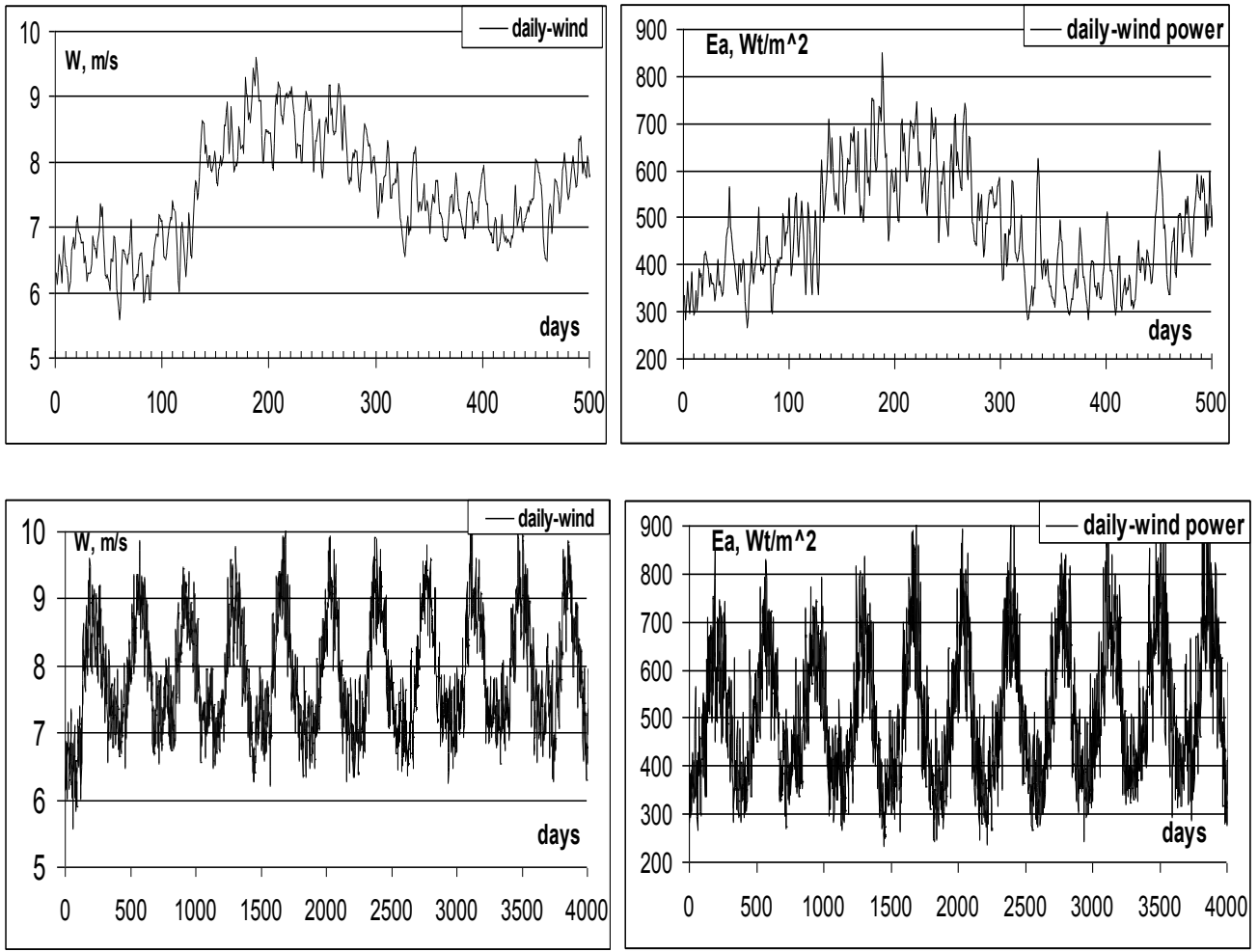

Fig. 4-app. The time history of wind and wind power averaged for a day and whole IO/
Left panels: detailed (upper) and full history (lower) series.
Right panels: the same but for wind power.

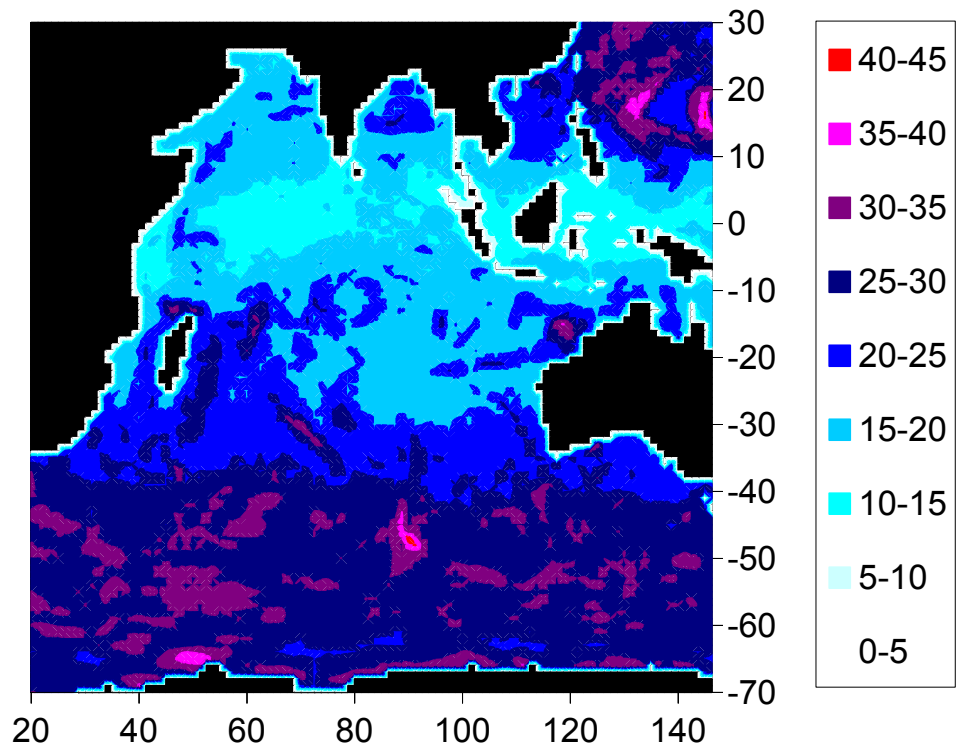

Fig. 5-app. The chart of maximal wind values found at each point of IO for the period 1998-2009 yy

**References**

[1] ftp://polar.ncep.noaa.gov/pub/history/waves/

[2] *Tolman H.L., and D.V. Chalikov.* Source terms in a third generation wind wave model// J. Phys. Oceanogr. 1996, V**26**, #11. P. 2497-2518.

[3] *Gulev S.K., Grigirieva V.* Last century changes in ocean wind waves height from global visual wave data// Geophys. Res. Lett. 2004, V 31, L24302, doi:10.1029/2004GL021040.

[4] Handbook for a wind and wave regime in the Bering and White seas/ Russian Shipping Registr. Lopatukhin L.I., Bukhanovkii A.V., Chernyshova E.S., eds. 2010. 565p.

[5] *Caires S. and Sterl A.* 100-Year Return Value Estimates for Ocean Wind Speed and Significant Wave Height from the ERA-40 Data// J. Climate. 2005, V.18, #4. P. 1032-1048.

[6] *Bauer E.* Characteristic frequency distribution of remotely sensed *in situ* and modeled wind speed// Int. J. Climatology. 1996. V.16. P. 1087-1102.

[7] *Monahan A.H.* The Probability Distribution of Sea Surface Wind Speeds. Part I: Theory andSeaWinds Observations//J. Climate. 2006, V.19, #2. P. 497-520

[8] *Monahan A.H.* The Probability Distribution of Sea Surface Wind Speeds. Part II: Dataset Intercomparison and Seasonal Variability//J. Climate. 2006, V.19, #2. P. 521-634

[9] *The WAMDI Group.* The WAM – a third generation ocean wave prediction model// J. Phys. Oceanogr., 1988. V. 18, p. 1775-1810.

[10] *Polnikov V.G., V. Innocentini.* Comparative study performance of wind wave model: WAVEWATCH- modified by the new source function. Engineering Applications of Computational Fluid Mechanics. 2008. V. 2, #4. P. 466-481

[11] *Polnikov V.G.* The role of wind waves in the dynamics of the air-sea interface // Izvestiya, Atmospheric and Oceanic Physics, 2009, Vol. 45, No. 3, pp. 346–356(Eng. Transl).

[12] *Per Kållberg, Paul Berrisford, Brian Hoskins, Adrian Simmons, Sakari Uppala, Sylvie Lamy-Thépaut and Rob Hine*. ERA-40 Atlas/ ERA-40 Project Report Series No.19. June 2005. http://www.ecmwf.int/publications/library/ecpublications/_pdf/era/era40/ERA40_PRS19_rev.pdf

[13] *Polnikov V.G.* Features of the adaptive spectral analysis for hydrodynamic processes // Proceedings of the 5$^{th}$ all-USSR meeting for statistical meteorology. Kazan'. 1985. Pp. 146-151 (in Russian) .

[14] *Obukhov A. M..* Turbulence and dynamics of the atmosphere. //Leningrad: Hydrometeoizdat Pablishing House. 1988.416 p (in Russian).